\documentclass[aps,prc,twocolumn,showpacs,superscriptaddress,floatfix,10pt]{revtex4-1}
\usepackage{graphicx}% Include figure files
\usepackage{dcolumn}% Align table columns on decimal point
\usepackage{bm}% bold math
\usepackage{amsfonts}
\usepackage{amssymb}
\usepackage{amsmath}
\begin{document}

% a few definitions
\newcommand{\vcr}{\mbox{${\bf r}\,$}}

\title{Energy dependence of potential barriers and its effect on fusion cross-sections}

% Force line breaks with \\
\author{A.S. Umar}
\affiliation{Department of Physics and Astronomy, Vanderbilt University, Nashville, Tennessee 37235, USA}
\author{C. Simenel}
\affiliation{Department of Nuclear Physics, RSPE, Australian National University, Canberra, ACT 0200, Australia}
\author{V.E. Oberacker}
\affiliation{Department of Physics and Astronomy, Vanderbilt University, Nashville, Tennessee 37235, USA}

\date{\today}

%------------------------------------------------------------------------------
\begin{abstract}
\begin{description}
\item[Background]
%CS Modeling of experimental fusion cross-sections in terms of fusion barriers and the incorporation of quantum
%mechanical processes that take place during the onset of fusion is one of the challenging problems in low-energy nuclear
%reactions. Similarly,  experimentally obtained fusion barriers and barrier distributions provide an insight into the
%heavy-ion reaction dynamics leading to fusion for energies in the vicinity of the Coulomb barrier. Theoretically the
%coupled-channels method is the commonly used approach to study fusion barriers and barrier distributions by modeling the
%coupling of the relative motion to internal excitations of nuclei during the early stages of the collision. Recently, it
%was shown that the time-dependent Hartree-Fock theory can be used to provide the ingredients used in coupled-channels
%calculations.  This together with the recently developed density-constrained time-dependent Hartree-Fock theory for
%calculating fusion barriers and the corresponding fusion cross-sections opens the door to studying heavy-ion fusion from
%a microscopic standpoint.
Couplings between relative motion and internal structures are known to affect fusion
barriers by dynamically modifying the densities of the colliding nuclei. The effect is expected to be stronger at energies near
the barrier top, where changes in density have longer time to develop than at higher energies.
This gives rise to an energy dependence
of the barriers as predicted by modern time-dependent Hartree-Fock (TDHF) calculations [K. Washiyama and D. Lacroix,
Phys. Rev. C {\bf 78}, 024610 (2008)]. %, which are free of any parameter adjusted on reaction mechanisms.
Quantitatively, modern TDHF calculations are able to predict realistic fusion thresholds. However, the  evolution of the
potential barrier with bombarding energy remains to be confronted with the experimental data.
\item[Purpose] The aim is to find signatures of the energy dependence of the barrier by comparing fusion cross-sections
calculated from potentials obtained
at different bombarding energies with the experimental data.
\item[Method]  This comparison is made for the $^{40}$Ca+$^{40}$Ca and $^{16}$O+$^{208}$Pb systems.
Fusion cross-sections are computed from potentials calculated with the density-constrained TDHF method.
\item[Results] The couplings decrease the barrier at low-energy in both cases.
A deviation from the Woods-Saxon nuclear potential is also observed at the lowest energies.
%Sub-barrier fusion cross-sections are well reproduced by the lowest energy potential but overestimate above-barrier data.
In general, fusion cross-sections around a given energy are better reproduced by the potential calculated at this energy.
%CS We observe that nucleon transfer and collective excitations present in TDHF collisions manifests itself in terms
%of energy-dependent interaction barriers.
%Microscopically calculated fusion barriers are able to reproduce, for the most part,  both the gross and the detailed
%features of the experimental fusion cross-sections.
The coordinate-dependent mass plays a crucial role for the
reproduction of sub-barrier fusion cross-sections. % but does not affect above barrier results.
% At above barrier energies barrier peaks computed via direct TDHF
% calculations and the ones obtained using the density-constrained TDHF method scale close to each other.
Effects of the energy dependence of the potential can be found in experimental barrier distributions only if the
variation of the barrier is significant in the energy-range spanned by the distribution. It appears to be the case for
$^{16}$O+$^{208}$Pb but not for $^{40}$Ca+$^{40}$Ca.
\item[Conclusions] These results show that the energy dependence of the barrier predicted in TDHF calculations is
realistic. This confirms that the TDHF approach can be used to study the couplings between relative motion and internal
degrees of freedom in heavy-ion collisions.
% provides a reasonable description of the
%dynamics inherent in the early stages of low-energy heavy-ion collisions leading to fusion.
\end{description}
\end{abstract}
% PACS, the Physics and Astronomy Classification Scheme.
\maketitle

%------------------------------------------------------------------------------
\section{Introduction}

Experimentally obtained fusion cross-sections are generally interpreted in terms of models
involving a nucleus-nucleus potential %CS tunneling through a one-dimensional
barrier, which results from the combination of the attractive
nuclear force and the repulsive Coulomb interaction.
The reduction of the many-body fusion to a one-dimensional potential barrier problem
requires the isolation of the most important physical processes that contribute to the
building of the correct effective barrier.

Experimental fusion barrier distributions~\cite{rowley1991} obtained from
the low-energy fusion reactions of heavy-ions shed some light into the detailed microscopic mechanisms
that are in play during the entrance channel dynamics on the way to fusion~\cite{leigh1995,dasgupta1998}.
In particular, they may serve as a microscope to discern various inelastic excitations and transfer mechanisms
which couple to the relative motion.
This coupling to internal degrees of freedom induces a splitting~\cite{dasso1985} and/or a renormalization of the
barrier~\cite{hagino1997}.
%CSmay otherwise be obscured in the fusion excitation functions.
%CS Naturally, barrier distributions will also arise if one or both of the participating nuclei
%possess a static deformation. In this case, different barriers will be obtained as a function of the entrance
%channel alignments of the nuclei.  Since the entangling of barriers due to static deformation and those due to other
%excitation mechanism may be nontrivial it is advantageous to examine fusion reactions of spherical nuclei for the latter.
The primary underlying mechanism is the dynamical change in the density along the fusion path which modifies the potential energy.

Obviously, this density change is not instantaneous.  For instance, it was shown in Ref.~\cite{simenel2013b} that the
development of a neck due to couplings to  octupole phonons in $^{40}$Ca+$^{40}$Ca could take approximately one
zeptosecond.  As a consequence, the dynamical change of the density is most significant at low energy (near the barrier-top)
where the colliding partners spend enough time in the vicinity of each other with little relative kinetic energy.  At high
energies, however, the nuclei overcome the barrier essentially in their ground-state density.  This energy dependence of
the effect of the couplings on the density evolution was clearly shown in time-dependent Hartree-Fock (TDHF)
calculations by Washiyama and Lacroix for the same systems~\cite{washiyama2008}. This naturally translates into an energy
dependence of the nucleus-nucleus potential, similar to what was introduced phenomenologically in the Sao-Paulo
potential~\cite{chamon2002}.  Consequently, the barrier corresponding to near barrier-top energies includes dynamical
couplings effects
and can be referred to as a {\it dynamic-adiabatic} barrier, while at high energy the nucleus-nucleus interaction is
determined by a {\it sudden} potential which can be calculated assuming frozen ground-state densities.

Due to the dynamical nature of this effect time-dependent approaches are well suited for this study.  The
energy-dependence of the ion-ion potentials have been studied using several approaches based on the fully microscopic TDHF
theory~\cite{washiyama2008,oberacker2010}.  It is usually found that the barrier heights increase with bombarding
energy. However, this increase is quite slow as the sudden potential is recovered at typically twice (or more) the
barrier-top energy~\cite{washiyama2008}.

To date, the validity of the TDHF approach in describing the fusion mechanism for heavy-ions has essentially been tested by
comparing TDHF fusion thresholds with experimental barriers~\cite{washiyama2008,simenel2008,simenel2013b}, although few
fusion excitation functions from direct TDHF calculations~\cite{bonche1978,simenel2008,simenel2013a} and $\gamma$-decay
spectra~\cite{simenel2007} associated to pre-equilibrium giant-dipole resonance
\cite{simenel2001,simenel2007,oberacker2012,umar2005a} have also been compared with experimental data.  Nevertheless, the
agreement between experimental barriers and TDHF predictions is only for near barrier-top energies, i.e., for the
dynamic-adiabatic barrier. Indeed, the predicted transition from the dynamic-adiabatic barrier to the sudden barrier
with increasing energy remains to be validated by comparisons with experimental data.
The purpose of this study is to accomplish this goal.

Towards this goal, we have calculated fusion cross-sections for the
$^{40}$Ca+$^{40}$Ca and $^{16}$O+$^{208}$Pb systems using potentials associated with different bombarding energies.
These
potentials were computed using the density-constrained TDHF (DC-TDHF) method~\cite{umar2006b} using realistic TDHF
trajectories.  The comparison of the resulting fusion cross-sections with experimental data is used to identify
signatures for the energy dependence of the barrier.
%CS In addition, theoretical modeling of the fusion cross-sections and fusion barrier distributions  for the
%$^{16}$O+$^{208}$Pb system has so far been a challenge to standard coupled-channel
%calculations~\cite{morton1999,hinde2002,dasgupta2007,esbensen2007,hagino2007}.

In the next section we give a brief outline of the TDHF and DC-TDHF methods used in the calculations.
This is followed by the calculation of barriers and fusion cross-sections for the $^{40}$Ca+$^{40}$Ca system
and subsequently the $^{16}$O+$^{208}$Pb system.
The paper ends with the summary and conclusions that can be drawn from the results.

\section{Theoretical Outline}

\subsection{Theoretical tools to describe fusion}

Theoretically, the coupled-channels (CC) method is the most commonly used approach to study fusion
barriers (see Ref.~\cite{hagino2012} for a review).
The standard CC approach for calculating
heavy-ion fusion cross sections contains several adjustable parameters
which determine the bare nucleus-nucleus potential which is often
assumed to be of Woods-Saxon form. These potential parameters are
usually fitted to measured fusion cross sections or to elastic
scattering data. In addition, experimental data
such as energies and B(E$\lambda$) values of collective vibrations
and giant resonances are required as input for the CC calculations
to determine the collective coupling potentials.
This is a limitation for exotic nuclei for which these data are not always available.
%CS and hence the predictive power of the CC method may be limited in such cases.
A possible solution of this problem is to compute these parameters directly with
microscopic models~\cite{umar2006b,washiyama2008,simenel2013b} and use them in standard
coupled channel calculations~\cite{simenel2013b}.
Finally, it is  difficult to incorporate multi-nucleon transfer
channels %and dynamical neck formation
into the CC formalism.
%CS Ultimately, one would like to have an approach for calculating internuclear potentials which is time-dependent and
%is unrestricted in the choice of collective variables.

Alternatively, fully microscopic theories could be used to overcome these limitations.
In particular, they only require an effective interaction or an energy-density functional to
describe the interactions between the nucleons.
Of course, microscopic approaches are much more time-consuming from a computational point of view and one has to consider
approximations to the exact quantum many-body problem.
The theoretical formalism for the
microscopic description of complex many-body quantum dynamics
and the understanding of the nuclear interactions %that result in self-bound, composite nuclei possessing the observed properties
are the underlying challenges for studying low energy nuclear reactions.

\subsection{Time-dependent Hartree-Fock method}

The time-dependent Hartree-Fock (TDHF)  theory is a mean-field approximation of the exact time-dependent many-body problem.
It  provides a good starting point for a
fully microscopic theory of large amplitude collective
motion~\cite{negele1982,simenel2012} including fusion reactions.
But only in recent years has it become feasible to perform TDHF calculations on a
three-dimensional Cartesian grid without any symmetry restrictions
and with accurate numerical methods~\cite{reinhard1988,umar1991a,kim1997,maruhn2005,nakatsukasa2005,umar2006c,guo2008}.
In addition, the quality of energy-density functionals has been substantially
improved~\cite{chabanat1998a,kluepfel2009,kortelainen2010}.
One limitation of the TDHF approach is that it can only be used for fusion at above barrier energies
since the theory does not allow for many-body tunneling.
Nevertheless, the TDHF fusion threshold provides a prediction of the dynamic-adiabatic
barrier-top energy in a very good agreement with experimental
data~\cite{simenel2008,washiyama2008}.
The TDHF theory has then been used to study
%Both the coupled-channel approach~\cite{hagino2012} and TDHF calculations have been used to study
 the couplings
between fusion and collective excitations such as rotational motion~\cite{simenel2004,umar2006d,umar2008b} and vibrational modes
\cite{simenel2001,simenel2007,oberacker2012,simenel2013a,simenel2013b}.
%CS Recently, it was shown that a microscopic approach like TDHF could be used to determine the parameters entering
%coupled channel calculations~\cite{simenel2013b}. %The purpose of this paper is to use the time-dependent microscopic
%approach to examine fusion barriers for

%The Hartree-Fock approximation
%and its time-dependent generalization %CS the time-dependent Hartree-Fock theory
%have provided possible means to study the diverse phenomena
%observed in low energy nuclear physics~\cite{negele1982,simenel2012}.

Given a many-body Hamiltonian $\hat{H}$, %CS containing two and three-body interactions
%\begin{equation}
%H=\sum_i^N t_i + \sum_{i<j}^Nv_{ij}+\sum_{i<j<k}^Nv_{ijk}\;,
%\end{equation}
the  action $S$ can be constructed as
\begin{equation}
S=\int_{t_1}^{t_2}dt<\Phi(t)|\hat{H}-i\hbar\partial_t|\Phi(t)>\;.
\end{equation}
Here, $\Phi$ denotes the time-dependent correlated many-body
wavefunction, $\Phi({\bf r_1,r_2,\ldots,r_{A}};t)$. %CS, and $t_i$ is the one-body kinetic energy operator.
The variational principle $\delta S=0$ is then equivalent to
%CS General variation of $S$ recovers
the time-dependent Schr\"odinger equation.
In the TDHF approximation the many-body wavefunction is replaced by a single
Slater determinant and this form is preserved at all times.
The determinental form guarantees the antisymmetry required by the Pauli
principle for a system of fermions. In this limit, the
variation of the action yields the most probable time-dependent mean-field path
between points $t_1$ and $t_2$ in the multi-dimensional
space-time phase space:
\begin{equation}
\delta S=0 \rightarrow \Phi_0(t)\;,
\label{variat}
\end{equation}
where  $\Phi_0(t)$ is  a Slater determinant
with the associated single-particle states $\phi_{\lambda}(\vcr,t)$.
The variation in Eq.(\ref{variat}) is performed with respect to
the single-particle states $\phi_{\lambda}$ and $\phi^{*}_{\lambda}$. % we obtain a
This leads to a set of coupled, nonlinear, self-consistent initial value equations
for the single-particle states
\begin{equation}
h\left( \left\{ \phi_{\mu} \right\} \right) \phi_{\lambda}=i\hbar
\dot{\phi_{\lambda}}
\;\;\;\;\;\;\;\;\;\lambda=1,...,N\;,
\label{tdhf0}
\end{equation}
and their Hermitic conjugates.
These are the fully microscopic TDHF equations.
%CS which preserve the major conservation laws such as the particle number, total energy, total angular momentum, etc.
As we see from Eq.(\ref{tdhf0}), each single-particle state evolves in the
mean-field generated by the concerted action of all the other single-particle
states.

In standard TDHF applications to heavy-ion collisions, the initial nuclei are calculated using the static Hartree-Fock (HF) theory and the Skyrme
functional~\cite{chabanat1998a}. The resulting Slater
determinants for each nucleus comprise the larger Slater determinant describing the colliding
system during the TDHF evolution.
Nuclei are assumed
to move on a pure Coulomb trajectory until the initial separation between the nuclear centers used
in TDHF evolution.
Of course, no assumption is made on the subsequent  trajectory in the TDHF evolution.
Using the Coulomb trajectory we compute the relative kinetic energy at this
separation and the associated translational momenta for each nucleus. The nuclei are then boosted
by multiplying the HF states with
\begin{equation}
\Phi _{j}\rightarrow \exp (\imath\mathbf{k}_{j}\cdot \mathbf{R})\Phi _{j}\;,
\end{equation}
where $\Phi _{j}$ is the HF state for nucleus $j$ and $\mathbf{R}$ is the corresponding
center of mass coordinate
\begin{equation}
\mathbf{R}=\frac{1}{A_{j}}\sum _{i=1}^{A_{j}}\mathbf{r}_{i}\;.
\end{equation}
The Galilean invariance and the conservation of the total energy in the Skyrme TDHF equations are used to check the
convergence of the calculations.

Due to the fact that TDHF calculations do not include sub-barrier tunneling of the many-body wave-function,
the fusion probability, $P_{fus.}(L,E_{\mathrm{c.m.}})$, for a particular
orbital angular momentum $L$ at the center-of-mass energy $E_{\mathrm{c.m.}}$ can only be
$P_{fus.}^{TDHF}=0$ or~$1$.
As a consequence the quantal expression for the fusion cross-section
\begin{equation}
\sigma_{fus.}(E_{\mathrm{c.m.}}) = \frac{\pi\hbar^2}{2\mu E_{\mathrm{c.m.}}} \sum_{L=0}^\infty (2L+1) P_{fus.}(L,E_{\mathrm{c.m.}})\;,
\label{eq:cs}
\end{equation}
where $\mu$ is the reduced mass of the system, reduces to
\begin{eqnarray}
\sigma_{fus.}(E_{\mathrm{c.m.}}) &=& \frac{\pi\hbar^2}{2\mu E_{\mathrm{c.m.}}} \sum_{L=0}^{L_{max}(E_{\mathrm{c.m.}})} (2L+1) \nonumber \\
&=&  \frac{\pi\hbar^2}{2\mu E_{\mathrm{c.m.}}} [L_{max}(E_{\mathrm{c.m.}})+1]^2\;,
\end{eqnarray}
$L_{max}$ being the largest  orbital angular momentum leading to fusion.
This is known as the quantum sharp cut-off formula~\cite{blair1954}.

Since TDHF is based on the independent-particle approximation
it can be interpreted as the semi-classical limit of a fully quantal theory thus allowing
a connection to macroscopic coordinates and providing insight about the collision process.
In this sense the TDHF dynamics can  only be used to compute
the semiclassical trajectories of the collective moments of the composite system as a
function of time. Note that the part of the residual interaction which is neglected in TDHF may produce
fluctuations and correlations which affect these trajectories.
Recent beyond TDHF developments have been used to investigate the effects of such fluctuations in heavy-ion
collisions~\cite{simenel2011,washiyama2009b}.
However, the TDHF approach is optimized to the expectation values of one-body operators~\cite{balian1981}
and is then capable to predict these quantities.
This was demonstrated by the recent successes of TDHF in reproducing various reaction mechanisms in heavy-ion collisions.
Moreover, beyond TDHF calculations remain numerically difficult.
We then restrict the present calculations to the TDHF level.

%\cite{kim97,sim01,uma06a,mar06,sim07,guo12,sim13b},

One of the main application of recent TDHF codes has been to study fusion reactions.
For TDHF collisions of light and medium mass systems, as well as highly mass-asymmetric systems,
fusion generally occurs immediately above the Coulomb barrier.
In heavier systems, however,
there is an energy range above the barrier where fusion does not occur~\cite{simenel2012,guo2012,umar2010a}.
This phenomenon is the microscopic analogue of the macroscopic \textit{extra-push}
threshold~\cite{swiatecki1982}.
In the extreme case of actinide collisions, fusion becomes impossible and the fragments reseparate in few zeptoseconds~\cite{tian2008,golabek2009,kedziora2010}.
%CS In the lower part of this energy range deep-inelastic collisions are dominant while at slightly higher energies the
%system develops a long-lived and pronounced neck reminiscent of a dinuclear configuration. The outcome of these
%long-lived configurations is uncertain due to the absence of quantum decay processes and transitions. For these systems
%TDHF results in a compact configuration only for energies considerably above the static potential energy surface.
%However, despite the high energy,

The path to fusion as described in TDHF calculations is a sequence of states from dinuclear configurations to a compact compound system.
Along this path, one-body dissipation plays a crucial role and
single-particle friction can quickly absorb the kinetic energy of the relative motion.
%CS and lead to a configuration that may be considered a thermal doorway state.
As long as the average single-particle excitation energy per nucleon %in this doorway state
is less than the shell energy (about $4-8$~MeV) the details of the ground state
potential energy surface are still felt and shell correction energies influence the
TDHF dynamics.
It is precisely for this reason that the DC-TDHF approach allows us to
reproduce ion-ion interaction barriers for heavy-ion collisions.

\subsection{DC-TDHF method}

The TDHF theory does not include quantum tunnelling of the many-body wave function.
Consequently, direct TDHF calculations cannot be used to describe sub-barrier fusion.
Nevertheless, a number of approaches based on TDHF were developed to extract fusion
potentials with dynamical effects~\cite{umar2006b,washiyama2008} in order to compute  fusion cross-sections at sub-barrier energies.

The density-constrained TDHF (DC-TDHF) utilizes a novel approach of using time-dependent densities from
TDHF to self-consistently calculate the underlying
ion-ion interaction potentials~\cite{umar2006b} and excitation energies~\cite{umar2009a}.
These potential barriers then allow for the calculation of fusion
cross-sections at both sub-barrier and above-barrier energies. The method was applied
to calculate fusion and capture cross sections above and below the barrier,
ranging from light systems %CSrelevant for the neutron star crust
~\cite{simenel2013a,umar2012a}
to hot and cold fusion reactions leading to superheavy element $Z=112$~\cite{umar2010a}.
In all cases a
good agreement between the measured fusion cross sections and the DC-TDHF results was found.
This is rather remarkable given the fact that the only input in TDHF is the
Skyrme energy-density functional whose parameters are determined from structure
information.
%CS and there are no adjustable parameters.

The concept of using density as a constraint for calculating collective states
from TDHF time-evolution was first introduced in Ref.~\cite{cusson1985}, and used
in calculating collective energy surfaces in connection with nuclear molecular
resonances in Ref.~\cite{umar1985}.
In this approach
the TDHF time-evolution takes place with no restrictions.
At certain times during the evolution the instantaneous density is used to
perform a static Hartree-Fock minimization while holding the neutron and proton densities constrained
to be the corresponding instantaneous TDHF densities. In essence, this provides us with the
TDHF dynamical path in relation to the multi-dimensional static energy surface
of the combined nuclear system. The advantages of this method in comparison to other mean-field
based microscopic methods such as the constrained Hartree-Fock (CHF) method are obvious. First,
there is no need to introduce external constraining operators which assume that the collective
motion is confined to the constrained phase space.
Second, the static adiabatic approximation is
replaced by the dynamical analogue where the most energetically favorable state is obtained
by including sudden rearrangements and the dynamical system does not have to move along the
valley of the potential energy surface. In short we have a self-organizing system which selects
its evolutionary path by itself following the microscopic dynamics.
All of the dynamical features included in TDHF are naturally included in the DC-TDHF calculations.
These effects include neck formation, mass exchange,
internal excitations, deformation effects to all order, as well as the effect of nuclear alignment
for deformed systems.

In the DC-TDHF method the ion-ion interaction potential is given by
\begin{equation}
V_{DC}(R)=E_{\mathrm{DC}}(R)-E_{\mathrm{A_{1}}}-E_{\mathrm{A_{2}}}\;,
\label{eq:vr}
\end{equation}
where $E_{\mathrm{DC}}$ is the density-constrained energy at the instantaneous
separation $R(t)$, while $E_{\mathrm{A_{1}}}$ and $E_{\mathrm{A_{2}}}$ are the binding energies of
the two nuclei obtained with the same effective interaction.
This ion-ion potential $V_{DC}(R)$ is asymptotically correct
since at large initial separations it exactly reproduces $V_{Coulomb}(R_{max})$.
In addition to the ion-ion potential it is also possible to obtain coordinate
dependent mass parameters. One can compute the ``effective mass'' $M(R)$
using the conservation of energy
\begin{equation}
M(R)=\frac{2[E_{\mathrm{c.m.}}-V_{DC}(R)]}{\dot{R}^{2}}\;,
\label{eq:mr}
\end{equation}
where the collective velocity $\dot{R}$ is directly obtained from the TDHF evolution.
This coordinate dependent mass can be exactly
incorporated into the ion-ion potential, which we call $V(R)$, by using a point-transformation~\cite{goeke1983,umar2009b}.
The effect of the coordinate-dependent mass is to modify the inner part of the ion-ion potential,
which is important for fusion cross-sections at deep sub-barrier energies.

Fusion cross-sections are calculated by directly integrating the Schr\"odinger equation
\begin{equation}
\left [ \frac{-\hbar^2}{2\mu} \frac{d^2}{d{R}^2} + \frac{\hbar^2 \ell (\ell+1)}{2 \mu {R}^2} + V({R})
 - E_\mathrm{c.m.} \right] \psi_{\ell}({R}) = 0 \;,
\label{eq:Schroed1}
\end{equation}
using the well-established {\it Incoming Wave Boundary Condition} (IWBC) method~\cite{hagino1999} to obtain
the barrier penetrabilities $P_{fus.}(L,E_{\mathrm{c.m.}})$ which determine the total fusion cross
section
[Eq.~(\ref{eq:cs})].

In writing Eq.~(\ref{eq:vr}) we have introduced the concept of an adiabatic reference state for
a given TDHF configuration.
%The difference between the energy of this state, $E_{DC}[R]$,
%and the total TDHF energy represents the internal energy.
The adiabatic reference state is the one obtained via the density constraint calculation.
It is the Slater determinant with lowest energy for the given density with vanishing current.
It is then used to approximate the collective potential energy~\cite{cusson1985}.
We would like to
emphasize  that this procedure does not affect the TDHF time-evolution and
contains no free parameters or normalization.

Finally, ion-ion interaction potentials calculated using DC-TDHF correspond to the
configuration attained during a particular TDHF collision. For light and
medium mass systems as well as heavier systems for which fusion is
the dominant reaction product, DC-TDHF calculations at near barrier-top energy
give a fusion barrier which is expected to match the TDHF fusion threshold.
In practice, due to the underlying numerical approximations in the DC-TDHF method,
small (typically less than 0.5~MeV) underestimation of the TDHF fusion threshold
are sometime observed.

%CS with an appreciable but relatively small energy dependence. On the other hand, for reactions leading to superheavy
%systems fusion is not the dominant channel at barrier-top energies. Instead the system sticks in some dinuclear
%configuration with possible break-up after exchanging a few nucleons. The long-time evolution to break-up is beyond the
%scope of TDHF due to the absence of quantum decay processes and transitions. As we increase the energy above the barrier
%this phenomenon gradually changes to the formation of a truly composite object. This is somewhat similar to the
%extra-push phenomenon discussed in phenomenological models.

\section{Results}

TDHF calculations for the DC-TDHF computation of microscopic potential barriers for the
$^{40}$Ca+$^{40}$Ca system were done in a Cartesian box which is $50$~fm along the collision axis and $25$~fm in
the other two directions. The nuclei were placed at an initial separation of $20$~fm.
For the $^{16}$O+$^{208}$Pb system we have chosen a
Cartesian box which is $60$~fm along the collision axis and $30$~fm in
the other two directions. The two nuclei are placed at an initial separation of $24$~fm.
Calculations used the SLy4 Skyrmefunctional~\cite{chabanat1998a} as described in Ref.~\cite{umar2006c}.
Static calculations are done using the damped-relaxation method~\cite{bottcher1989}.
The numerical accuracy of the static binding energies and the deviation of the computed DC-TDHF potential
from the point Coulomb energy in the initial state of the collision dynamics is of the order
of $50-150$~keV. We have performed density constraint calculations at every $10-20$~fm/c interval.

\subsection{$^{40}$Ca+$^{40}$Ca Fusion Barriers}
Recently, particular experimental attention has been given to fusion reactions involving
Ca isotopes~\cite{stefanini2009,jiang2010,montagnoli2010,montagnoli2012}.
These new experiments supplement older fusion data~\cite{aljuwair1984} and extend them to lower sub-barrier energies.
In Ref.~\cite{montagnoli2012} a comprehensive CC calculation for this system has also been presented utilizing
the \textit{shallow potential} approach~\cite{esbensen2007}. These calculations use M3Y+repulsion potential
and the excitations of collective phonons.
In particular, octupole vibrations have been shown to play an important role on the dynamics in this system~\cite{rowley2010,simenel2013b}.
%CS $2_1^+$, $3^-$, $5^-$, the states arising from the deexcitation of the two-phonon quadrupole triplet, in total 24
%channels. Similarly, the influence of octupole photons for Calcium isotopes was suggested in Ref.~\cite{rowley2010}.

\subsubsection{Nucleus-nucleus potentials}

\begin{figure}[!htb]
\includegraphics*[width=8.6cm]{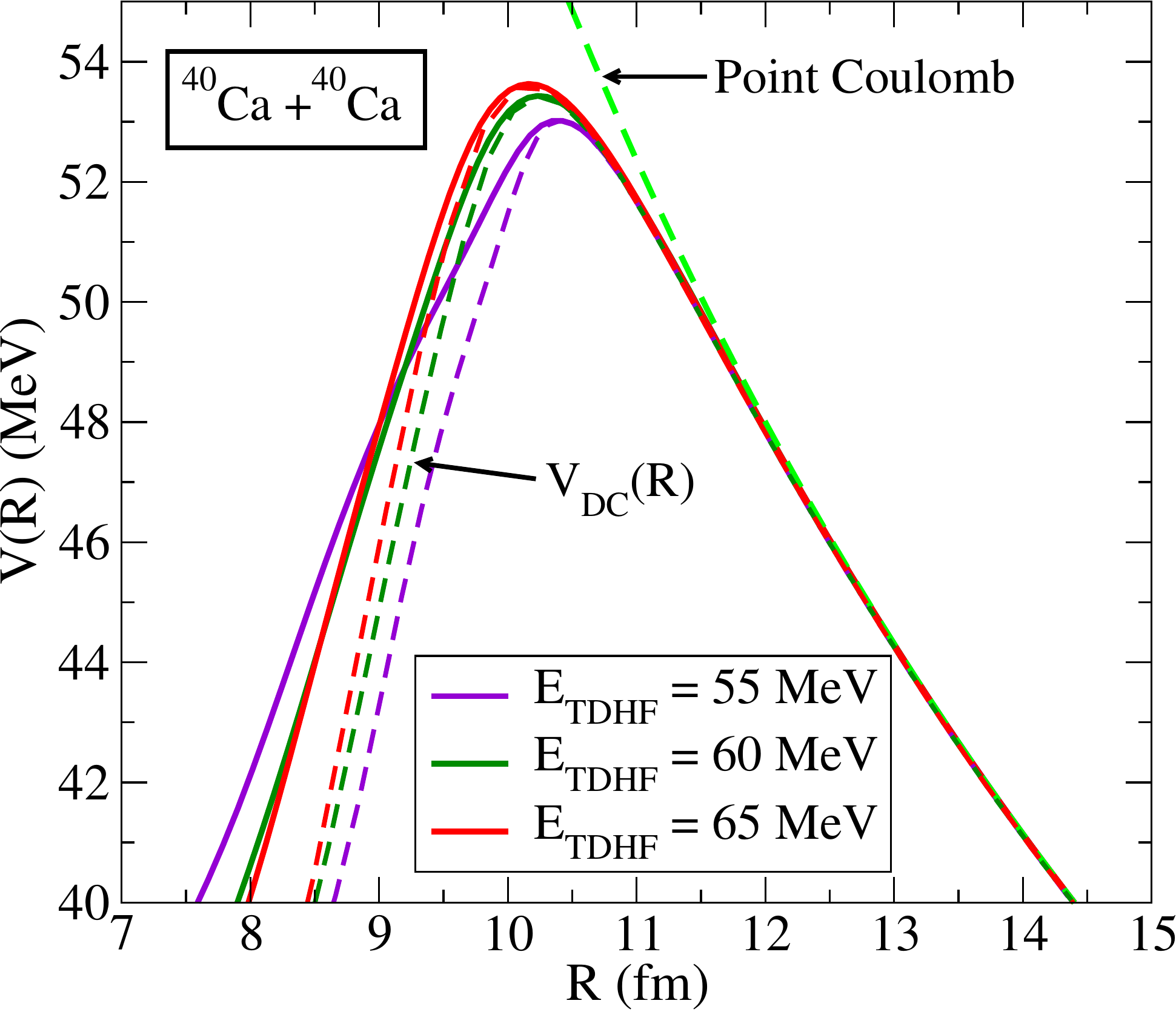}
\caption{\protect (Color online) DC-TDHF ion-ion interaction potentials $V(R)$ (solid lines) including coordinate
dependence of the effective mass $M(R)$ for $^{40}$Ca+$^{40}$Ca obtained from TDHF
calculations at various center-of-mass energies.
The potentials $V_{DC}(R)$ obtained without the coordinate dependence of $M(R)$ are plotted with thin dashed lines.
Shown also is the corresponding point-Coulomb potential (thick dashed line).}
\label{fig1}
\end{figure}

The $^{40}$Ca+$^{40}$Ca system was investigated in Ref.~\cite{keser2012} with the DC-TDHF method using TDHF c.m. energies $55$, $60$, and $65$~MeV.
The resulting potential barriers are reported in Fig.~\ref{fig1}.
In the present work, additional calculations have been performed to study in more details the energy dependence of the barrier and its effect on
the fusion cross-sections.
We have performed TDHF calculations in $1$~MeV intervals in the $53-65$~MeV range and computed the corresponding DC-TDHF potentials.
As a result, barrier heights are in the range of $52.6-53.6$~MeV all located in the vicinity
of nuclear separation $R=10.2$~fm.
We observe that for the $^{40}$Ca+$^{40}$Ca system DC-TDHF potential barriers do not show an apparent strong energy dependence.

In the DC-TDHF method the energy dependence of the barriers arises from the changing dynamical behavior of the system.
At energies close to the barrier-top the onset of neck dynamics is slow and allow ample time for density rearrangements
for the system, whereas as the energy is increased there is less and less time for rearrangements to
occur and a long-lived neck to form, thus approaching the frozen-density limit~\cite{washiyama2008}.
The barrier corresponding to the lowest TDHF energy may be called the \textit{dynamic-adiabatic} barrier
as opposed to a \textit{static-adiabatic} barrier that could be obtained by using the constrained Hartree-Fock
approach or a prescription like the folding-model. The barrier corresponding to TDHF energies much higher
than the dynamic-adiabatic barrier may be labeled as the \textit{sudden} barrier.
%CS (not frozen since a neck is formed even at high energies).
We see from Fig.~\ref{fig1} that this leads to an increasing barrier height with increasing collision energy
and quickly saturates for energies that are considerably higher than the lowest energy barrier.
In this sense, we obtain a distribution of barriers as a function of collision energy.
%CS but these barriers, while they include the time-dependent density at zero impact parameter, may not have a
%one-to-one correspondence to the fusion cross-sections directly obtained from a full TDHF calculation by finding the
%maximum impact parameter value for fusion. A full TDHF calculation at above barrier energies is a dynamical many-body
%collision that includes other dynamical effects not present in the reduced problem. In this sense the DC-TDHF method
%provides us with microscopic barriers between the dynamic-adiabatic and sudden regimes.

An important dynamical effect is due to the coordinate-dependence of the mass, $M(R)$.
In Fig.~\ref{fig1} this effect is demonstrated by plotting the direct DC-TDHF potentials, $V_{DC}(R)$
(dashed lines), and those that include the modification of the coordinate-dependent mass, $V(R)$ (solid lines).
For TDHF collisions of symmetric systems
the net particle transfer is zero and cannot affect $M(R)$. However, the
 dynamical neck formation and collective excitations
are possible and can change the effective mass.

The potentials shown in Fig.~\ref{fig1} should not be directly compared with nucleus-nucleus potentials entering CC calculations.
Indeed, the latter
%CSThis makes the direct comparison of DC-TDHF barriers and those obtained from CC calculations difficult since CC
%barriers have a prescribed analytic form, usually Wood-Saxon, and
are un-coupled potentials
with various couplings and particle transfer added on subsequently. In cases were double-folding method is used the densities
are frozen as the nuclear separation $R$ changes. This usually implies a higher un-coupled barrier
height as it was found to be in the range $54.1-54.7$~MeV in Refs.~\cite{washiyama2008,simenel2013b,misicu2011,montagnoli2012}.

\subsubsection{Fusion cross-sections}

\begin{figure}[!htb]
\includegraphics*[width=8.6cm]{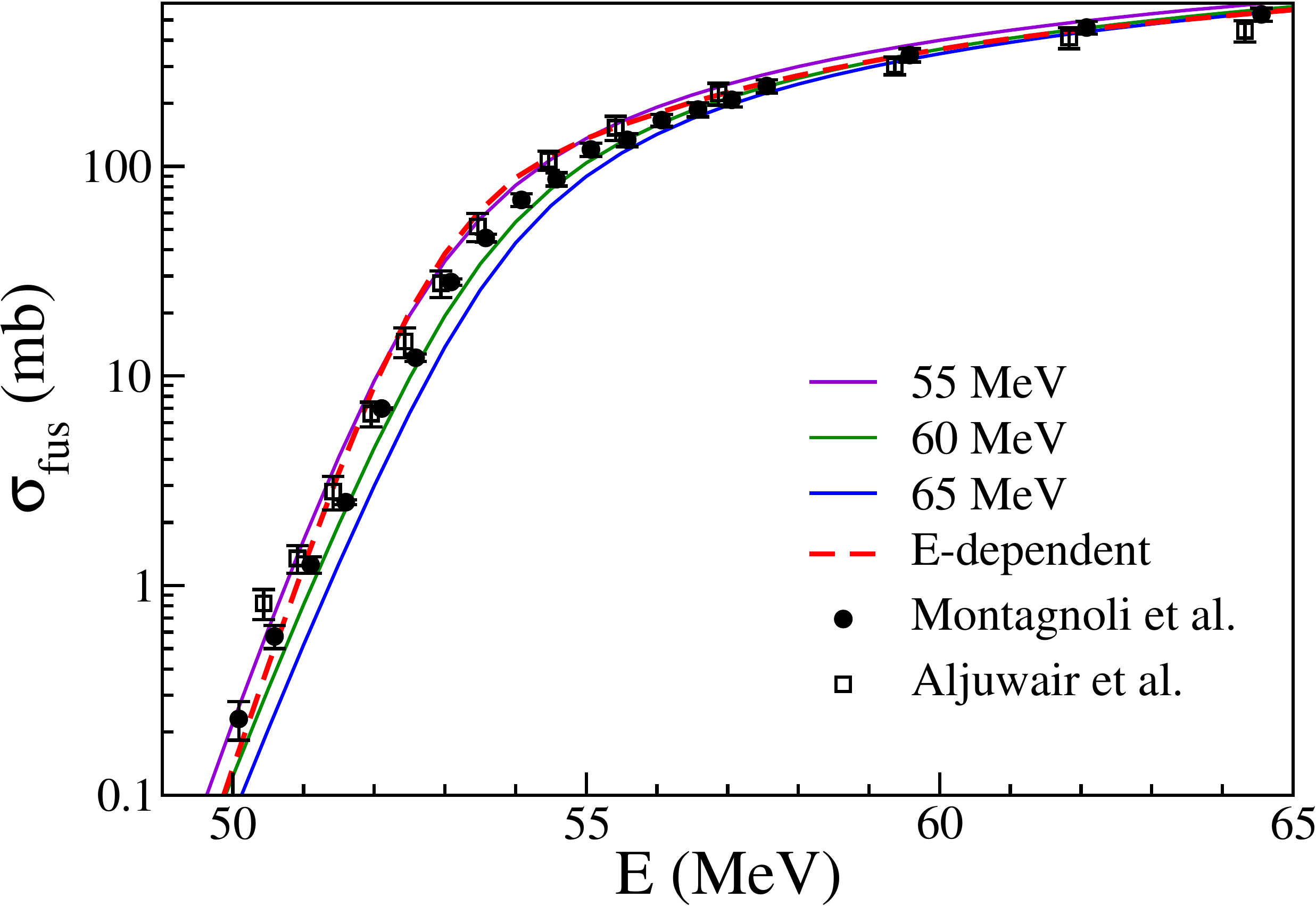}
\caption{\protect (Color online) Fusion cross-sections for $^{40}$Ca+$^{40}$Ca obtained from the DC-TDHF potentials
shown in Fig.~\ref{fig1}.  The dashed line represents the combined cross-sections $\bar{\sigma}(E)$.
The data points are from Refs.~\cite{montagnoli2012,aljuwair1984}.
\label{fig2}}
\end{figure}

\begin{figure}[!htb]
\includegraphics*[width=8.6cm]{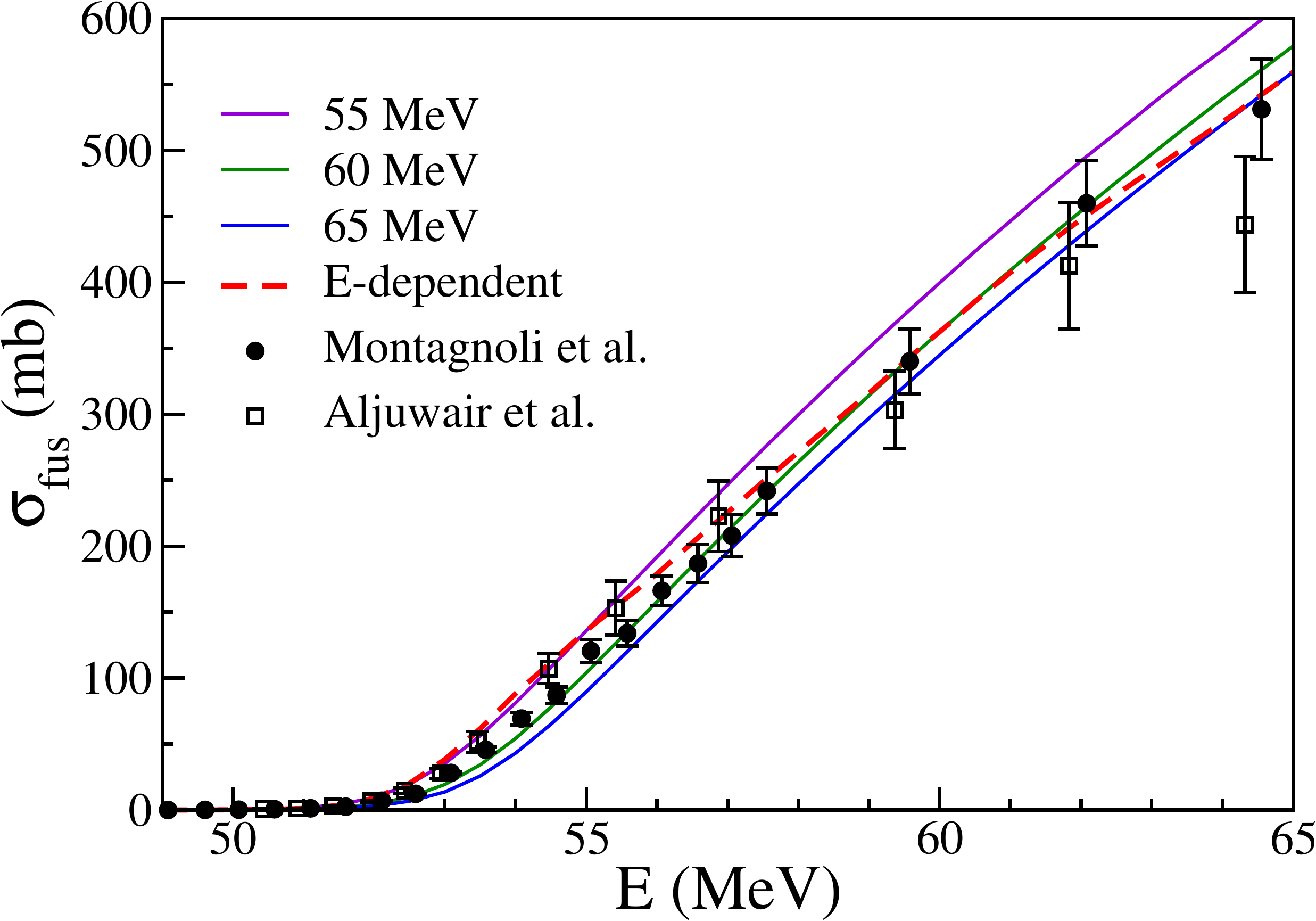}
\caption{\protect (Color online) Same as Fig.~\ref{fig2} in linear vertical scale.
\label{fig3}}
\end{figure}

%In Fig.\ref{fig2} we show
The corresponding fusion cross-sections calculated from the potentials $V(R)$ shown in Fig.~\ref{fig1} are plotted in Fig.~\ref{fig2} in logarithmic
scale and in Fig.~\ref{fig3} in linear scale.
%CS by integrating the Schr\"odinger equation Eq.~(\ref{eq:Schroed1}) using the IWBC method as described in the previous
%Section.
The experimental points are from Refs.~\cite{montagnoli2012,aljuwair1984}.
%CS We again observe the dependence of
The cross-sections clearly depend on the TDHF energy used to extract the DC-TDHF potential.
%CS The E-dependent cross-section shown in Fig.~\ref{fig2} is obtained by averaging the energy-dependent cross-sections.
%In order to gain further insight concerning the energy dependence of these cross-sections we also plot them on a linear
%scale in Fig.~\ref{fig3}. We make the following observations;
The interaction potential corresponding to the lowest TDHF energy leads to fusion cross-sections
which are in good agreement with the sub-barrier fusion data
but overestimate the cross-sections at higher energies.
On the other hand, the potential corresponding to the highest energy reproduces the highest energy data
but underestimates the data at lower energies.
%CS These two trends are what one might expect considering the physical arguments given above.
%CS Similarly, potentials corresponding to intermediate energies reproduce the data in that range.
%CS The largest disrepancy arises for energy regime slightly above the lowest barrier peak, although in this case the
%deviations are on the order of a few tens of $mb$.

In principle, each set of cross-sections $\sigma_n(E)$ is valid only near the TDHF energy $E_n$ used to calculate the potential.
One can then generate a unique function $\bar{\sigma}(E)=\sum_n\sigma_n(E)f_n(E)$ where $f_n(E)$ is a weighting function peaked at $E=E_n$.
In practice, $\bar{\sigma}(E)$ has been generated using
\begin{eqnarray}
f_0&=&\left\{
\begin{array}{ll}
1&E<E_0 \\
\cos^2[\frac{\pi}{2} \,\frac{E-E_0}{\Delta E}] \,\,\,\,& E_0\le E\le E_1 \\
0&E>E_1
\end{array}\right. \nonumber \\
f_{0<n<N}&=&\left\{
\begin{array}{ll}
0&E<E_{n-1} \\
\cos^2[\frac{\pi}{2} \,\frac{E-E_n}{\Delta E}] \,\,\,\,& E_{n-1}\le E\le E_{n+1} \\
0&E>E_{n+1}
\end{array}\right. \nonumber \\
f_{N}&=&\left\{
\begin{array}{ll}
0&E<E_{N-1} \\
\cos^2[\frac{\pi}{2} \,\frac{E-E_N}{\Delta E}] \,\,\,\,& E_{N-1}\le E\le E_{N} \\
1&E>E_{N}
\end{array}\right. .\nonumber
\end{eqnarray}
$E_0$ is the lowest TDHF energy at which fusion is observed and from which a potential can be extracted, while $E_N$ is the maximum TDHF energy
considered in this work.
$\Delta E$ is the constant energy step in the TDHF calculations.
(The generalization to non-constant $\Delta E$ is trivial.)

The resulting $\bar{\sigma}(E)$ is labelled ''$E$-dependent'' in Figs.~\ref{fig2} and \ref{fig3}.
Considering the experimental error bars and the fluctuations between the data sets, we see that there is an overall
agreement between $\bar{\sigma}(E)$ and the experimental fusion cross-sections, despite a slight overestimation of the
more recent data from Ref.~\cite{montagnoli2012} in the barrier region.   It is then reassuring to observe that the
energy-dependent DC-TDHF potentials lead to reasonable reproduction of the data in the energy-range of the TDHF energy
used for their calculation.  This comparison with experimental data also confirms that the potential barrier ''seen'' by
the system at high energy is effectively higher than the one at low-energy. It is unfortunate, however, that this
energy-dependence cannot be investigated below the barrier. This is due to the fact that the TDHF calculations at
sub-barrier energies do not lead to fusion and, then, the DC-TDHF method cannot be applied to extract the potential in
this energy regime. Nevertheless, the good agreement between sub-barrier data and the theoretical cross-sections
calculated with the dynamic-adiabatic potential indicates that this energy dependence is likely to be small.

\subsubsection{Fusion barrier distributions}

To investigate possible signatures of the energy dependence at energies close to the barrier,
we have  calculated the following quantity~\cite{dasgupta1998}
\begin{eqnarray}
\label{eq:bd}
 D(E_i)&=&\left[\frac{d^{2}(E\sigma_{fus}(E))}{dE^{2}}\right]_i \\ \nonumber
       &\simeq&\left(\frac{(E\sigma)_{i+1}-2(E\sigma)_i-(E\sigma)_{i-1}}{\Delta E^2}\right)\;,
\end{eqnarray}
which is known as the fusion barrier distribution~\cite{rowley1991}.
It is essentially zero except in the energy range of the barrier and has then been widely used to study the effect of the couplings between
relative motion and internal structures on fusion barriers.
% using the formula for equally spaced energy points separated by $\Delta E$
%CS Due to the multiplication of the fusion cross-sections with energy $E$ and the second-derivative deviations of the
%calculated cross-sections from the data are amplified in the barrier distribution calculations.
As it was discussed in some detail in Ref.~\cite{dasgupta1998} the calculation of the barrier distribution using
the above formula is sensitive to the value of the energy separation $\Delta E$ used in the finite-difference
formula. Commonly, a value between $\Delta E=1-2$~MeV is used.
%CS Small values for $\Delta E$ show more structure whereas larger values do more smoothing.
%In Fig.~\ref{fig4} we show the

\begin{figure}[!htb]
\includegraphics*[width=8.6cm]{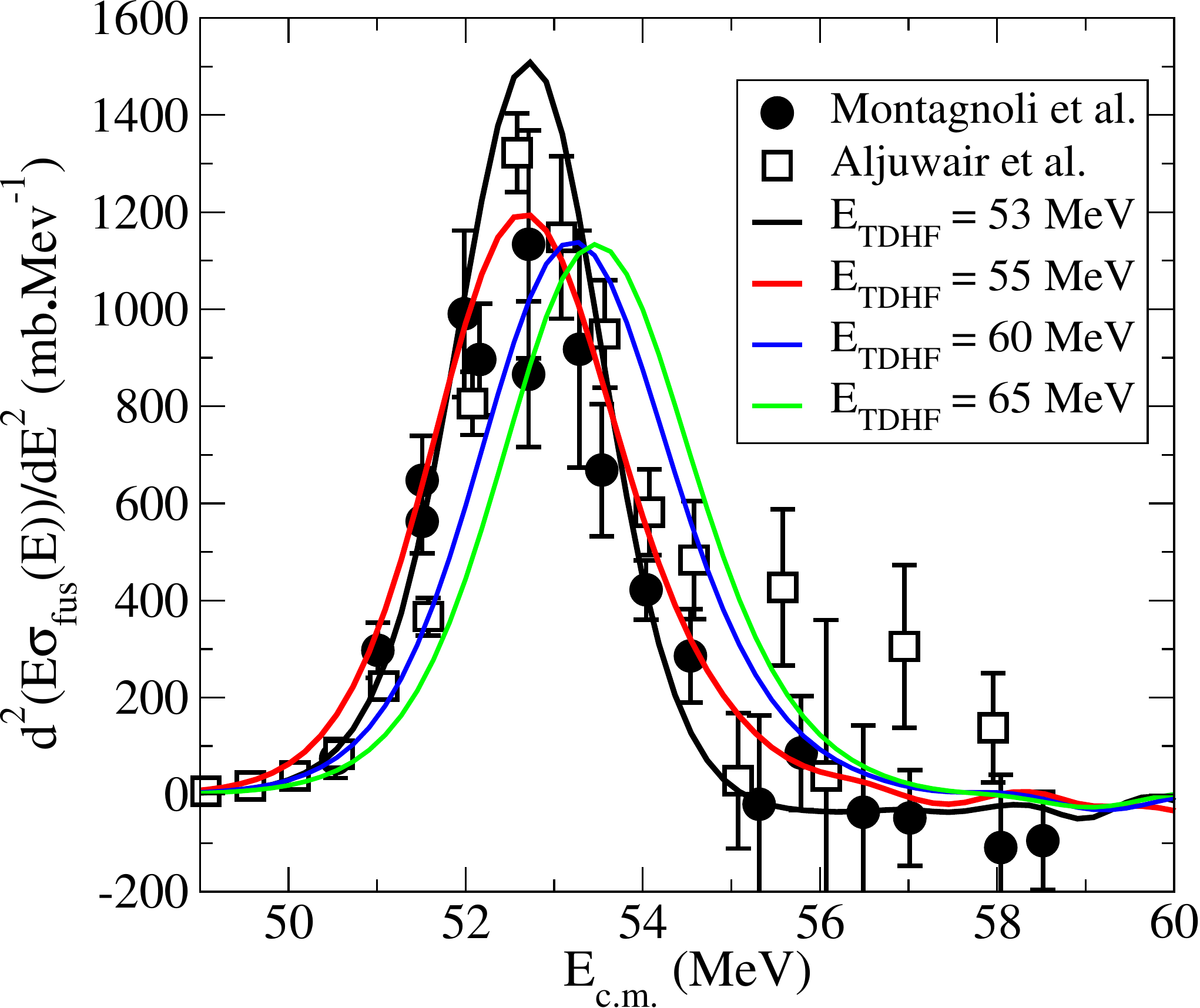}
\caption{\protect (Color online) Fusion barrier distributions for $^{40}$Ca+$^{40}$Ca obtained from DC-TDHF potentials
calculated with different TDHF energies.
%shown in Fig.~\ref{fig1}.
The data shown as solid-filled circles are from Ref.~\cite{montagnoli2012}, and the squares
are from Ref.~\cite{aljuwair1984}.
\label{fig4}}
\end{figure}

Selected barrier distributions obtained from different TDHF energies are shown in Fig.~\ref{fig4} together with experimental data from
Refs.~\cite{montagnoli2012,aljuwair1984}.
%CS calculated for each of the cross-section curve calculated from the three different TDHF energies shown in
%Fig.~\ref{fig2} as well as the lowest energy barrier corresponding to $53$~MeV. We also show %the recent data with solid
%circles~\cite{montagnoli2012} and the older data with squares~\cite{aljuwair1984}.
The barrier distributions were calculated with $\Delta E = 1$~MeV. % in accordance with the
%experimental curve~\cite{montagnoli2012}.
The distributions corresponding to different TDHF energies are generally smooth but the centroids
shift to a higher energy with increasing TDHF energy and the heights of the distributions become lower.
This change can be interpreted as being due to the difference in the dynamical processes that are more prevalent
at barrier-top energies in comparison to higher energies where we approach the frozen-density limit.
Despite fluctuations in the experimental data, it is clear that the distributions associated with the high TDHF energies ($E_{TDHF}=60$ and 65~MeV)
do not reproduce the experimental barrier distribution.
This is of course not a problem as the comparison should be made at energies close to 60-65~MeV, for which $D(E)\simeq0$.
Nevertheless, this indicates that the measured barrier distributions provide information on the dynamic-adiabatic barrier, but not on the potential
seen by the system at higher energies.

\subsection{$^{16}$O+$^{208}$Pb Fusion Barriers}

The second system we have studied is  $^{16}$O+$^{208}$Pb.  The choice of this system is partly motivated by the fact
that its fusion barrier is affected by early charge equilibration dynamics~\cite{simenel2008,simenel2010}. Quantitative
reproductions of fusion cross-sections for this system would then be an indication that the TDHF approach is able to
treat the interplay between nucleon transfer and fusion.  This system is also one for which fusion hindrance at deep
sub-barrier energies has been observed~\cite{dasgupta2007}. Standard coupled-channels calculations including low lying
vibrational states and one-neutron transfer channels  could not consistently reproduce the high and low-energy fusion
data. While the shallow-potential approach of Ref.~\cite{esbensen2007} had some success in reproducing the low-energy
part of the data it required an  imaginary potential to reproduce the high-energy part of the data. Furthermore,
inconsistencies in the shallow-potential approach for simultaneously reproducing the low and high-energy fusion data was
pointed out in Ref.~\cite{dasgupta2007}.

\subsubsection{Nucleus-nucleus potentials}

\begin{figure}[!htb]
\includegraphics*[width=8.6cm]{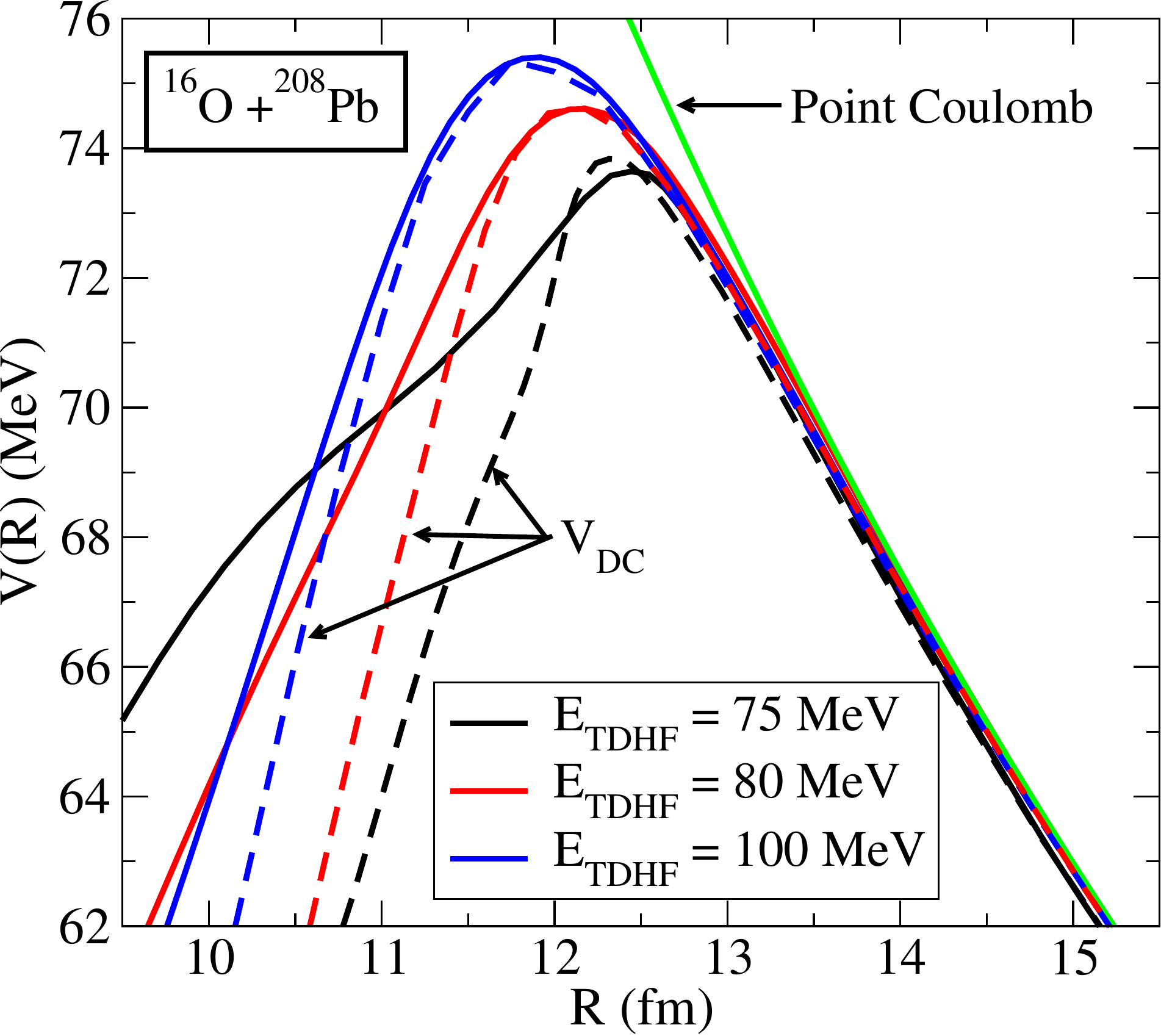}
\caption{
\protect (Color online) DC-TDHF ion-ion interaction potential $V(R)$ (solid lines) including coordinate dependence of
the effective mass $M(R)$ for $^{16}$O+$^{208}$Pb obtained from TDHF
calculations at various center-of-mass energies.
The potentials $V_{DC}(R)$ obtained without the coordinate dependence of $M(R)$ are plotted with thin dashed lines.
Shown also is the corresponding point-Coulomb potential.}
\label{fig5}
\end{figure}

We have performed TDHF calculations in $1$~MeV intervals between 75 and 80 MeV center-of-mass energies, as well
as at 90 and 100 MeV.
The corresponding DC-TDHF barriers are shown in Fig.~\ref{fig5} for $E_{TDHF}=75$, 80, and 100~MeV.
Barrier heights are in the range of $73.7-75.0$~MeV all located in the vicinity of nuclear separation $R=12$~fm.
As it was in the previous study the barrier thickness at sub-barrier energies changes with changing collision
energy due to the fact that at lower energies the system has more time to rearrange its density, which would
manifest itself as the formation of a neck followed by nucleon transfer~\cite{simenel2008,washiyama2008,umar2008a}
and collective excitations.
Similarly, the energy of the barrier-top is highest for highest energy approaching the sudden limit at
high energies.
Moreover, we observe that as we move down from the potential peak the inner part of the barrier
usually deviates from the Wood-Saxon+Coulomb form, which is the case for deep sub-barrier energies, with
or without the coordinate-dependent mass.

\begin{figure}[!htb]
\includegraphics*[width=8.6cm]{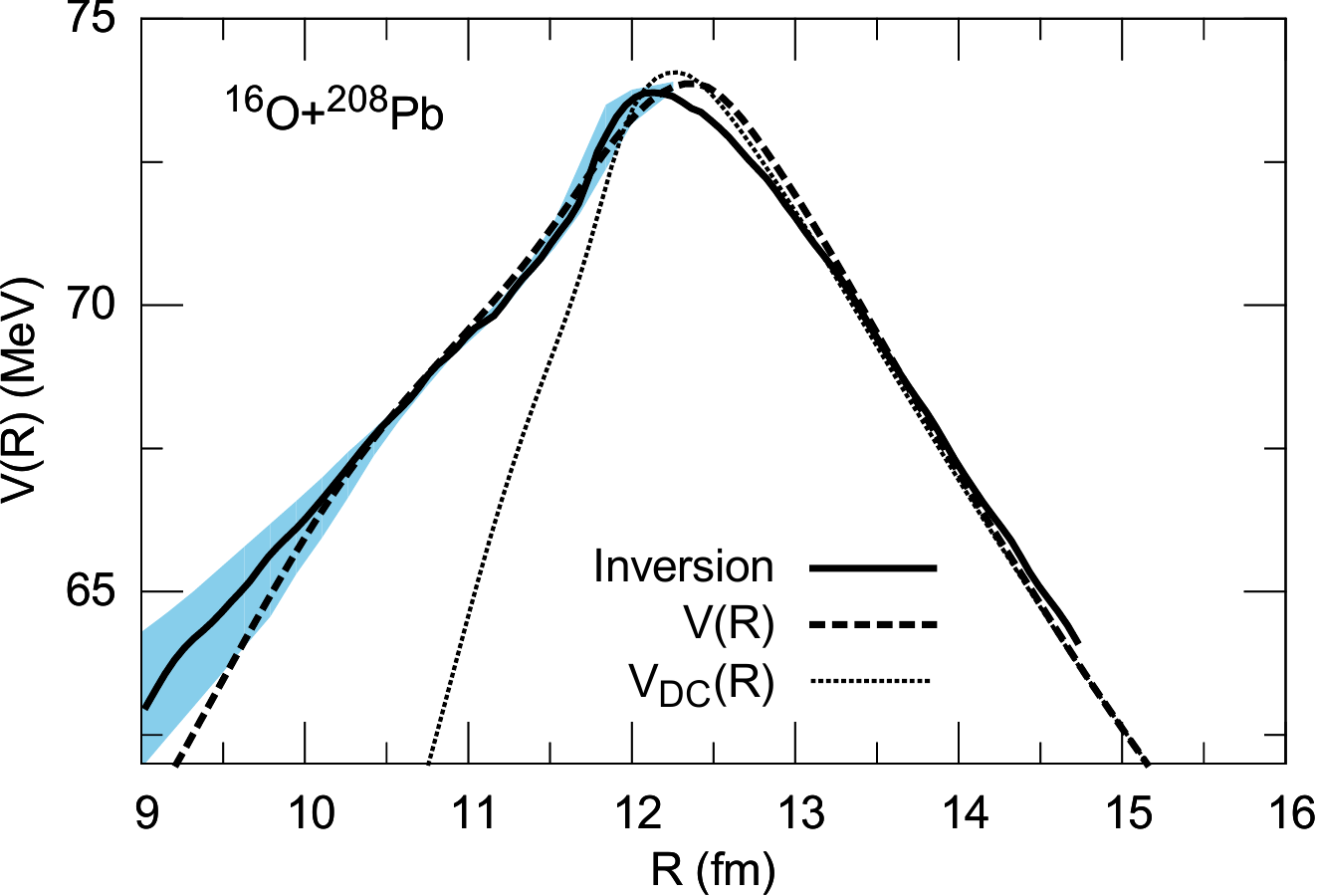}
\caption{(Color online) The \textit{adiabatic} potential obtained in Ref.~\cite{hagino2007} compared with the DC-TDHF
potential reproducing the sub-barrier cross-sections.}
\label{fig6}
\end{figure}

In Ref.~\cite{hagino2007} a method was developed to extract the ion-ion potential
directly from the experimental sub-barrier cross-sections in an attempt to understand the reason for CC
calculations not reproducing sub-barrier and high-energy part of the data with a single potential model.
These calculations showed that the form of the
potential deviated from the Wood-Saxon shape and one of the possible reasons to account for this deviation
was suggested to be the coordinate-dependent mass.
The potential barrier extracted directly from the sub-barrier data was called the \textit{adiabatic}
potential and is plotted in Fig.~\ref{fig6} (solid line, the shaded region indicates uncertainty)
together with the DC-TDHF potential at 75~MeV with (dashed line) and without (dotted line)  the coordinate-dependent mass.
The potential with coordinate dependent mass  is in much better agreement with the one extracted from data
using the inversion method.
We can conclude from these calculations that indeed the coordinate-dependent mass, which is really a
byproduct of heavy-ion and neck dynamics, is largely responsible for the thickening of the barrier for
deep sub-barrier energies as shown in Fig.~\ref{fig5}.

\subsubsection{Fusion cross-sections}

\begin{figure}[!htb]
\includegraphics*[width=8.6cm]{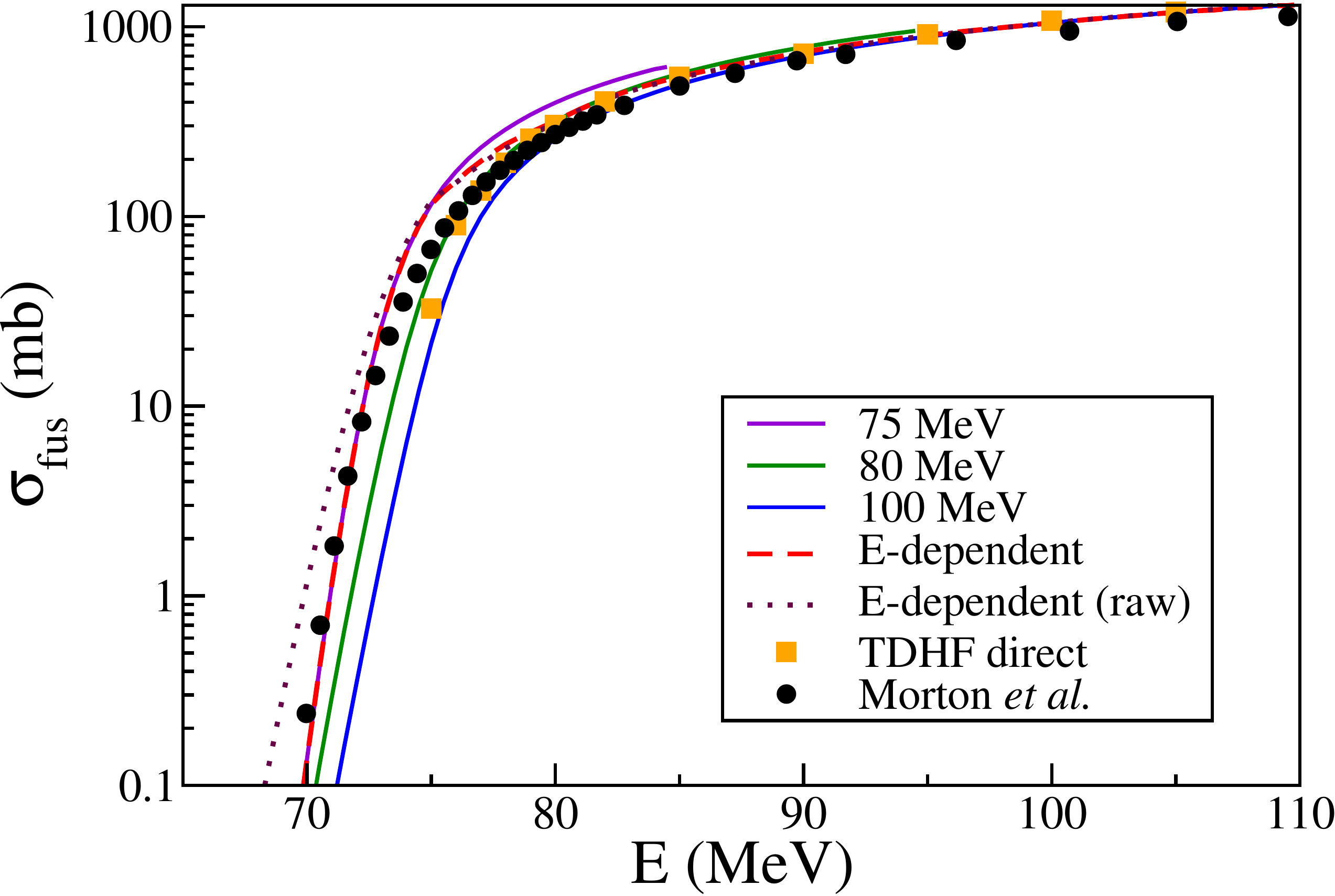}
\caption{\protect (Color online) Fusion cross-sections for $^{16}$O+$^{208}$Pb obtained from the DC-TDHF potentials
shown in Fig.~\ref{fig5}.  The dashed and dotted lines represent the combined cross-sections $\bar{\sigma}(E)$ obtained
with and without coordinate dependent mass, respectively. The data points are from Ref.~\cite{morton1999}.}
\label{fig7}
\end{figure}
\begin{figure}[!htb]
\includegraphics*[width=8.6cm]{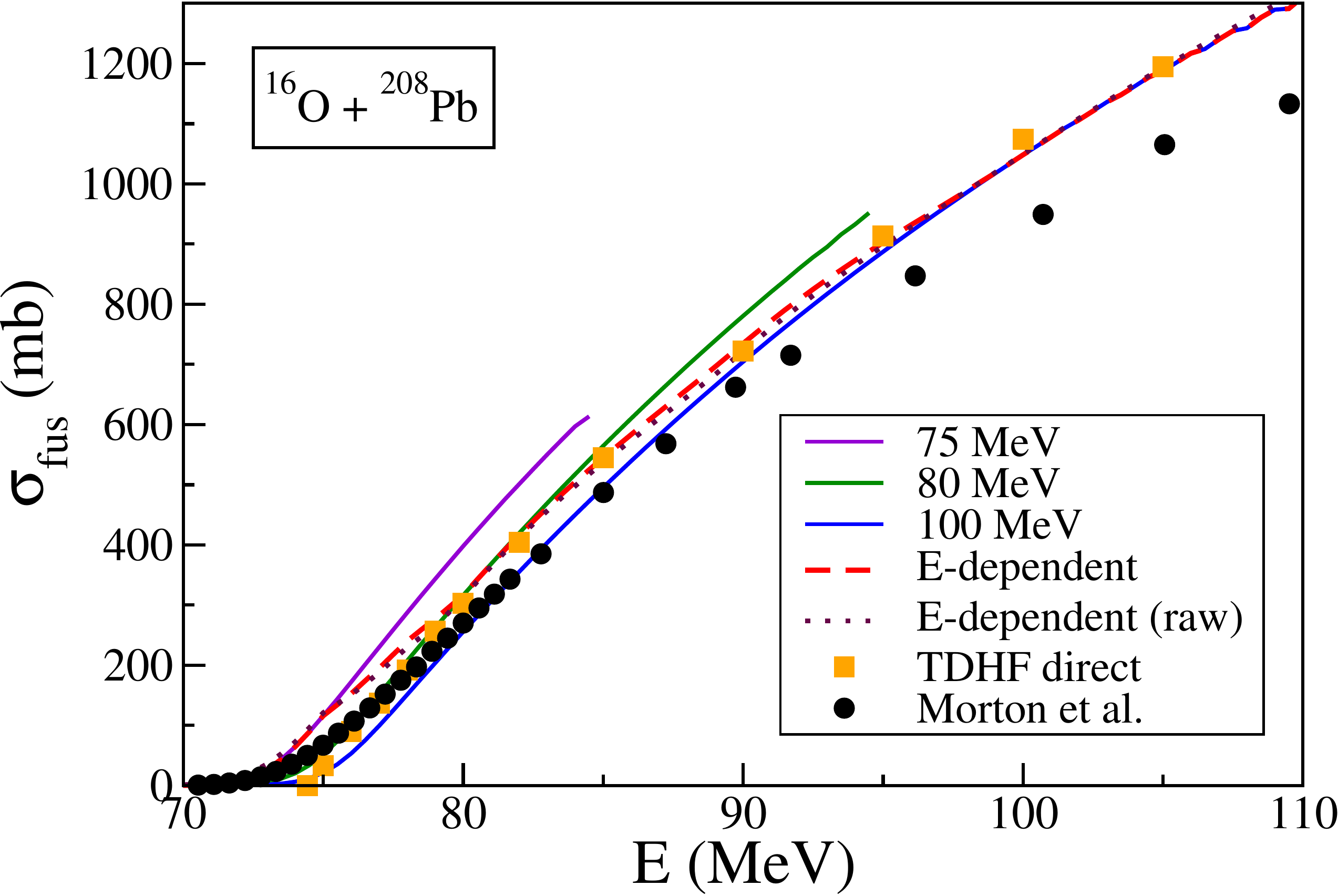}
\caption{\protect (Color online) Same as Fig.~\ref{fig7} in linear vertical scale.
\label{fig8}}
\end{figure}

As in the $^{40}$Ca+$^{40}$Ca reaction, the energy dependence of the ion-ion interaction potentials observed in Fig.~\ref{fig5} leads to the
corresponding change in the calculated
fusion cross-sections %CS, calculated from Eq.~(\ref{eq:Schroed1}) using the IWBC method,
as shown in Figs.~\ref{fig7} (logarithmic scale) and \ref{fig8} (linear scale).
Also shown in the same figures are the experimental cross-sections from Refs.~\cite{morton1999}.
The general trends observed in the energy-dependence of the cross-sections %shown in Fig.~\ref{fig7}
is similar
to the $^{40}$Ca+$^{40}$Ca case. % studied in the previous subsection.
%CS The lowest energy barrier reproduces the sub-barrier data quite well.
%CS For a more detailed examination of the energy-dependent cross-sections we plot them on a linear scale in Fig.~\ref{fig6}.
%The general trend is similar to the $^{40}$Ca+$^{40}$Ca case;
Indeed, the potential
obtained at the lowest TDHF energy reproduces the sub-barrier cross-sections.
In addition, the experimental cross-sections at high energy are better reproduced by potentials calculated at similar energies.
%CS while the intermediate and high energy barriers do not seem to do as well albeit for a parameter free theory 5-10\%
%error in fusion cross-sections may not be too unreasonable.
However, it is noticeable that, even in the energy range where they are supposed to be valid, the energy-dependent potential overestimates the experimental data.
Note that this is not a drawback of the method used to extract the potential as the problem can be traced back to the TDHF approximation itself.
Indeed, direct TDHF cross-sections computed at above barrier energies by finding the maximum impact parameter for fusion
at each energy overestimate the experimental cross-sections by the same amount (see Fig.~\ref{fig8}).  As the TDHF
calculations reproduce well the centroid of the experimental barrier distribution~\cite{simenel2008,washiyama2008}, it
is then likely that beyond mean-field effects are responsible for the observed discrepancy above the barrier. For
instance, the transfer of a proton pair, and, to a lesser extent, of an $\alpha$-cluster, which are not included in TDHF
calculations, have been shown to be an important mechanism in this system~\cite{videbaek1977,evers2011}.  Nevertheless,
this discrepancy is relatively small considering the fact that there are no free parameters.

%Note also that a similar overestimation of above-barrier fusion cross-sections has been observed in the
%$^{16}$O+$^{16}$O system~\cite{simenel2013a}. %This could indicate that the problem is in the description of the
%structure of $^{16}$O at the mean-field level.

%We observe that the direct TDHF results show similar differences from data for this system.
%This will be discussed further in the next section.
Let us now investigate the effect of the coordinate dependence of the effective mass $M(R)$ on the fusion cross-sections.
In Figs.~\ref{fig7} and \ref{fig8} we have also plotted the calculated E-dependent cross-sections without the use of the
coordinate-dependent mass (dotted curve).
Figure~\ref{fig8} shows that, above the barrier, the inclusion of the coordinate dependence of the effective mass does not play an important role as
both energy-dependent calculations lead to similar cross-sections.
However, Fig.~\ref{fig7} shows that this is not the case below the barrier.
Here, the effect of $M(R)$ on the low-energy cross-section is seen to be essential.
Indeed, without the coordinate dependence of the mass, the cross-sections are overestimated below the barrier.
Including this dependence widens the barrier (see Fig.~\ref{fig5}) and consequently reduces the cross-sections, providing a much better agreement
with the data (see Fig.~\ref{fig7}).

\subsubsection{Fusion barrier distributions}

Finally, we study the effect of the energy dependence of the potential on the fusion barrier distribution $D(E)$.
So far, standard CC calculations have not been able to reproduce the fusion barrier distributions consistently
for low and high energies~\cite{morton1999}.
%CS These calculations included excitation of single-phonon $3_1^-$ and $5_1^-$ states in $^{208}$Pb and $3_1^-$ state
%in $^{16}$O, as well as coupling to two-phonon states in $^{208}$Pb and particle transfer channels~\cite{morton1999}.
Improved barrier distributions at lower energies were calculated using CC
with the shallow-potential method~\cite{esbensen2007}.
%CS including the above excitations for target and projectile plus excitation of the $2_1^+$ state of $^{16}$O and
%excitation of $2_1^+$ and $4_1^+$ of $^{208}$Pb with a single neutron transfer.
However, above barrier cross-sections could only be explained with addition
of an imaginary absorbing potential.
%CS The centroid of the barrier distribution function considerably overestimated the data. In Ref.~\cite{hinde2002}
%inhibition of fusion due to mass-asymmetry for the $^{16}$O+$^{208}$Pb system was measured by comparing it with more
%mass symmetric systems leading to the same compound nucleus, whereas in Ref.~\cite{dasgupta2007} the need for gradual
%onset of quantum decoherence was proposed.
\begin{figure}[!htb]
\includegraphics*[width=8.6cm]{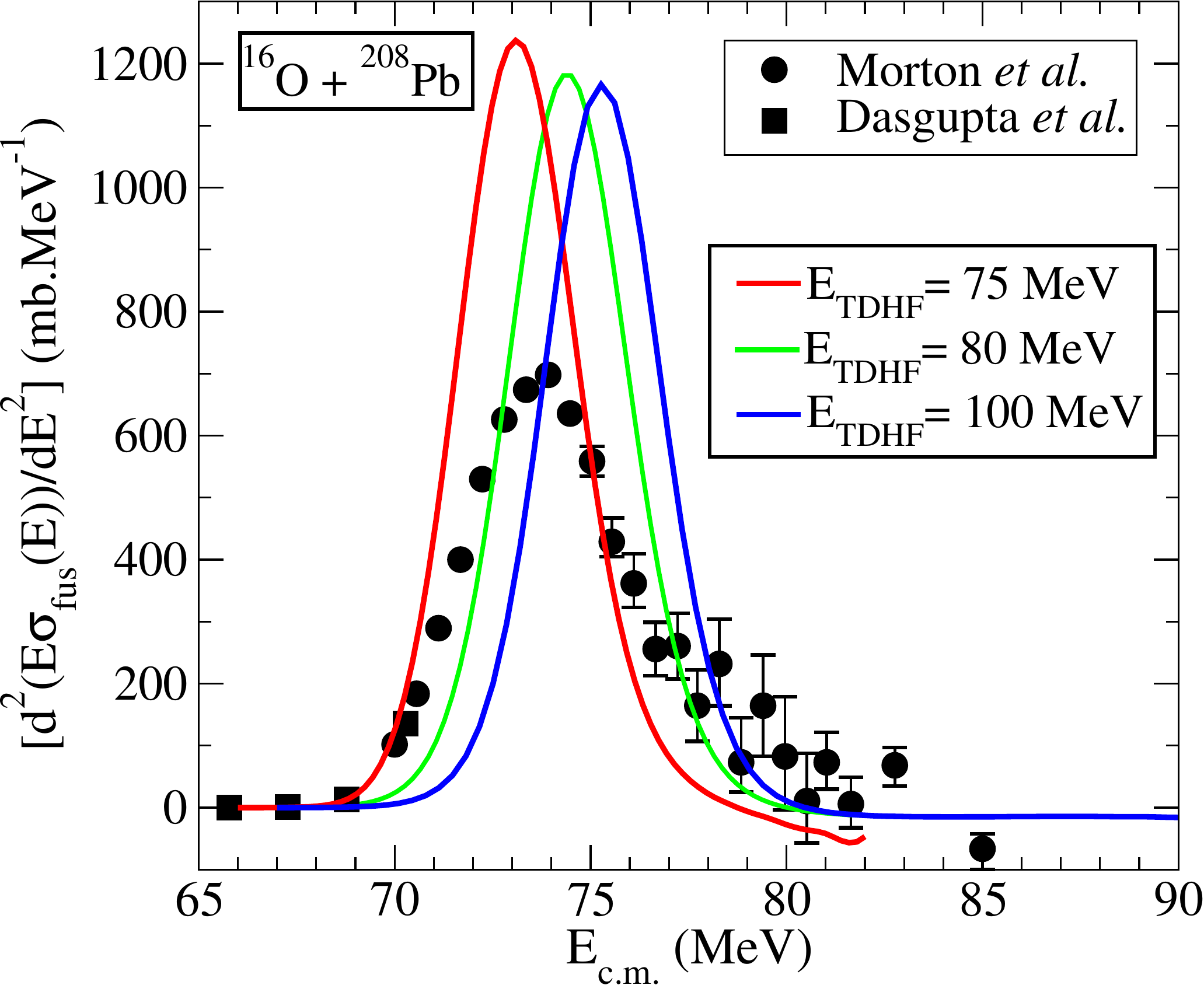}
\caption{\protect (Color online) Fusion barrier distributions for $^{16}$O+$^{208}$Pb obtained for various
TDHF energies. The data shown as solid-filled circles are from Refs.~\cite{morton1999,dasgupta2007}}.
\label{fig9}
\end{figure}
We have constructed fusion-barrier distributions from the DC-TDHF cross-section by using
an energy spacing of $\Delta E =2.0$~MeV as shown in Fig.~\ref{fig9} for TDHF bombarding energies 75, 80, and 100~MeV.
Also shown are the data from Refs.~\cite{morton1999,dasgupta2007}.
A first observation is that  the DC-TDHF barrier distributions suffer from the overestimation of the barrier-distributions at intermediate energies.
The difficulty in reproducing this region is shared with standard CC approaches.
The origin of this discrepancy are still unclear.
%CS Calculated barrier distributions at the lowest TDHF energy reproduce the low-energy part of the experimental
%distribution relatively well. On a log scale the barrier corresponding to the lowest TDHF energy goes over all the
%points in the 65$-$70~MeV range.
Another observation is that, as the TDHF energy is increased, the corresponding distributions peak at higher energies.
This qualitative observation was also made in the $^{40}$Ca+$^{40}$Ca case.
In addition, each theoretical barrier distribution is narrower than the experimental one.
This observation could be attributed to the energy dependence of the potential.
Indeed, the high energy tail of the experimental barrier distribution extends up to $\sim80$~MeV.
The barrier distribution computed from the $E_{TDHF}=75$~MeV potential naturally fails to reproduce the high energy part of the experimental barrier distribution.
The latter is much better reproduced by the $E_{TDHF}=80$~MeV potential.
The tail in the $75-80$~MeV region can then be interpreted as an effect of the gradual increase of the barrier height in this energy range.
Note that this effect is not visible in the $^{40}$Ca+$^{40}$Ca data due to the fact that the change in barrier height is not noticeable in the
limited energy range span by the barrier distribution.

%Similarly to CC calculations, these results suffer from the overestimation of the barrier-distributions at intermediate
%energies.  %CS  stemming from the overestimation of the fusion cross-sections in the same region. %CS While the dilemma
%of accurately reproducing the fusion barrier distributions for the $^{16}$O+$^{208}$Pb system persists the quality of
%the results obtained from DC-TDHF calculations in comparison to standard CC method is encouraging.

\section{Summary and discussion}

%CS We have also computed barrier heights (lowest collision energy for fusion) and fusion cross-sections for above
%barrier energies using direct TDHF calculations.  %For the fusion of  nuclei,

Ion-ion potentials are sensitive to the excitation and transfer mechanism in play on the way to forming a compound
nucleus. However, these couplings between relative motion and internal degrees of freedom have time to affect the
nucleus-nucleus potential only at low energy, leading to a ``dynamic-adiabatic'' potential.  At high energy, the system
does not have enough time to rearrange its density, leading to a ``sudden'' potential.  As a result, this leads to an
energy dependence of the potential and, in particular, of its barrier.  The purpose of this work was to identify
signatures of this energy dependence in experimental fusion cross-sections by comparing with the predictions of
microscopic calculations.

Fusion potentials around the barrier have been calculated for the $^{40}$Ca+$^{40}$Ca and $^{16}$O+$^{208}$Pb systems
using the DC-TDHF method based on TDHF density evolutions.  It is shown that, as we go to above barrier energies, the
energy dependence of the potential increases the barrier height and consequently slows down the increase of the fusion
cross-sections with increasing bombarding energy.  This effect happens in a large energy range until the sudden
potential is reached (according to Ref.~\cite{washiyama2008}, this can occur at about twice the energy of the barrier).
As a result, the dynamic-adiabatic and the sudden barriers can be very different.  The former reproduces sub-barrier
data, while the latter provides a better agreement at well above barrier energies than at low energies.   Discrepancies
remain, however, at above barrier energies for the $^{16}$O$+^{208}$Pb system, which could be due to proton-pair and
$\alpha$-cluster transfer not included in the theory.  It should also be noted that signatures of the energy-dependence
of the potential are less visible in the experimental barrier distributions due to the fact that these distributions
usually span a small energy range in which the variation of the barrier is not always very sensitive.

%CSWe observe that the dynamic-adiabatic barrier obtained at lowest TDHF energy does a very good job of reproducing the
%sub-barrier fusion data and reproduces the observed fusion hindrance in the  $^{16}$O+$^{208}$Pb system. %CS the
%accuracy of the comparison first suffers and subsequently improves for both systems. Similar behavior is found both in
%DC-TDHF and direct TDHF calculations. %This can be also seen the from

\begin{figure}[!htb]
\includegraphics*[width=8.6cm]{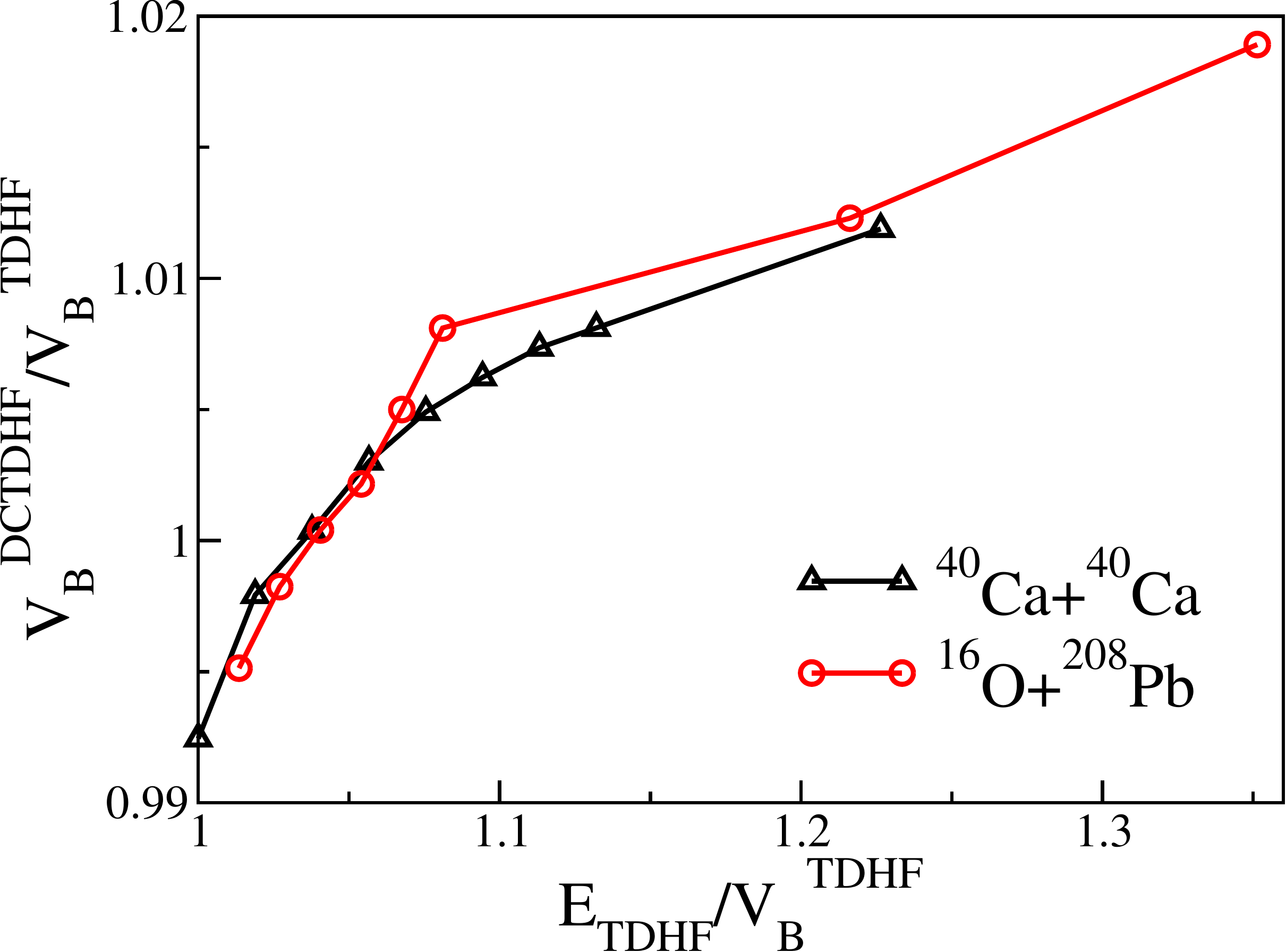}
\caption{(Color online) The ratio of barrier heights obtained from DC-TDHF and direct TDHF calculations
plotted against the dimensionless scale variable $E_{TDHF}/V_{B}^{TDHF}$.}
\label{fig10}
\end{figure}

Finally, let us compare the energy-dependence of the potentials in both systems. This is done in
Fig.~\ref{fig10} where we plot the ratio of the barrier heights
obtained from DC-TDHF, $V_B^{DC-TDHF}$, and direct TDHF, $V_B^{TDHF}$, calculations as a function of the dimensionless variable
$E_{TDHF}/V_{B}^{TDHF}$.
It is interesting to note that the energy dependence of the barriers are found to be very similar for both systems.
It is then not surprising that the same behavior is obtained in the fusion cross-section plots.
%CS This seems to suggest that certain reaction mechanisms may not be accounted for in this regime but improves again
%when we approach the sudden limit. On the other hand

The quality of the results suggests that the mean-field dynamics present in TDHF does properly account for
many of the excitation and transfer mechanisms.
Naturally, this is achieved in an average way as opposed to a fully quantal theory.
%CS The chosen systems may be considered ideal for static Hartree-Fock calculations as being doubly magic spherical
%nuclei. This is also the reason for their experimental preference since the excitation mechanisms are easily identified.
%Indeed, they are generally included in the fits of the Skyrme interaction. However, the Skyrme interaction does not
%contain any reaction input.
The present calculations are another testament to a growing number of TDHF calculations, both in the
small amplitude limit for low-lying and collective state calculations and in the large amplitude limit of reaction
dynamics, finding good comparisons with experimental observations.
This progress is partially due to the increased computational capabilities that allow such calculations to be
performed without using any symmetry restrictions and with modern energy density functionals.
This suggests that for low-energy heavy-ion
reactions TDHF remains as an ever more useful theoretical tool.

\begin{acknowledgments}
Useful discussions with M. Dasgupta and D. J. Hinde are acknowledged.
This work has been supported by the U.S. Department of Energy under Grant No.
DE-FG02-96ER40975 with Vanderbilt University,
and by the Australian Research Council under Grants No. FT120100760, No. FL110100098, and No. DP1094947.
Part of the calculations have been performed on the NCI National Facility in Canberra,
Australia, which is supported by the Australian Commonwealth Government.

\end{acknowledgments}

\bibliography{VU_bibtex_master.bib}

\end{document}